\documentstyle[multicol,pre,aps,epsf]{revtex}
\begin{document}

\title{The $\pm J $ spin glass in Migdal-Kadanoff approximation}

\author{Barbara Drossel$^1$ and M.A. Moore$^2$}
\address{${}^1$ School of Physics and Astronomy, Raymond and 
Beverley Sackler Faculty of Exact Sciences, Tel Aviv 
University, Tel Aviv 69978, Israel}
\address{${}^2$ Department of Physics, University of Manchester, 
Manchester M13 9PL, U.K.}
\date{\today} 
\maketitle

\begin{abstract}
We study the low-temperature phase of the three-dimensional $\pm J$
Ising spin glass in Migdal-Kadanoff approximation. At zero
temperature, $T=0$, the properties of the spin glass result from the
ground-state degeneracy and can be elucidated using scaling arguments
based on entropy. The approach to the asymptotic scaling regime is
very slow, and the correct exponents are only visible beyond system
sizes around 64.  At $T>0$, a crossover from the zero-temperature
behaviour to the behaviour expected from the droplet picture occurs 
at
length scales proportional to $T^{-d_s/2}$ where $d_s$ is the 
fractal
dimension of a domain wall.
Canonical droplet behaviour is not
visible at any temperature for systems whose linear dimension is
smaller than 16 lattice spacings, because
the data are either affected by the zero-temperature behaviour or the
critical point behaviour.
\end{abstract}

\begin{multicols}{2}
  
\section{Introduction}
\label{intro}

There is still no agreement about the nature of the low-temperature
phase of the Ising spin glass, which is defined by the Hamiltonian
$$H=-\sum_{\langle i,j\rangle} J_{ij} S_iS_j.$$ The spins can take 
the
values $\pm 1$, and the nearest-neighbour couplings $J_{ij}$ are
independent from each other and are most often chosen to be Gaussian
distributed with mean zero and a standard deviation $J$. 

While many Monte-Carlo simulations show properties conforming to the
replica-symmetry-breaking (RSB) scenario (implying  many
low-temperature states and a lack of self-averaging)
\cite{marinari,mprrz}, other simulations \cite{py} and analytical
arguments \cite{NS} favour the droplet picture (a scaling theory 
based
on the existence of only one low-temperature state and its time 
reverse).
The ambiguities 
stem from the difficulty in reaching the asymptotic limit of low
temperatures and large system sizes. Monte-Carlo results
are likely to be affected by finite-size and critical-point
effects. We have recently shown that a system that is known to
conform to the droplet picture at sufficiently large system sizes has
features similar to those of RSB if only small systems are studied 
and if the
temperature is not low enough \cite{ourprl,ourprb}. This system is 
the
hierarchical lattice, or, equivalently, the Migdal-Kadanoff
approximation (MKA) applied to a cubic or hypercubic lattice.  It is 
thus 
possible that  the Ising spin glass on three- or
four-dimensional lattices might show its true low-temperature 
properties
only beyond the length scales accessible to present-day Monte-Carlo
simulations.

Exact evaluation of ground states and low-lying excited states 
appears
to indicate a scenario that agrees neither with the droplet picture
nor with the RSB theory, but shows instead low-lying excitations 
which
are fractal \cite{PY00,KM00}. Newman and Stein have argued 
\cite{NS00}
that such excitations cannot exist in the thermodynamic limit. As the
studied system sizes are very small, the phenomenon might be a
small-size effect that vanishes at larger system sizes. Since fractal
excitations are not possible on a hierarchical lattice (only
combinations of compact droplets and domain walls can occur on it),
the MKA cannot show these low-lying excitations, and  agrees with the
droplet picture even for small system sizes at low temperatures with 
a 
Gaussian distribution for the bonds $J_{ij}$.

Very recently several papers have focussed on the
$\pm J$ Ising spin glass, where the nearest-neighbour couplings take
only the values 1 and $-1$, instead of being chosen from a Gaussian
distribution. Evidence is accumulating that the
ground-state degeneracy introduces new effects.  Thus, Krzakala and
Martin \cite{KM} argued that even if a system showed RSB at low
temperatures, different valleys in the energy landscape would differ
in entropy to the extend that for sufficiently large system sizes one
state would dominate the zero-temperature partition function, leading
for instance to a trivial overlap distribution (i.e. an overlap
distribution that is the sum of two $\delta$-functions at opposite
values of the overlap).  This argument is supported by simulations by
Palassini and Young \cite{PY} who find a crossover from a 
zero-temperature
behaviour with a trivial overlap distribution to a finite-temperature
behaviour which seems to agree with the RSB scenario. In contrast,
Hed, Hartmann and Domany, claim to find a non-trivial overlap
distribution even at zero temperature \cite{HHD}.

It is the purpose of this paper to study the low temperature
properties of the $\pm J$ model in MKA in order to shed some light on
the results of Monte-Carlo simulations, and to determine the
conditions under which the true low-temperature behaviour should be
visible. Our findings confirm the conjecture by Krzakala and Martin
that the zero-temperature behaviour is different from the
low-temperature behaviour, and they also confirm the scaling
assumptions concerning the entropy differences used in their
argument. Furthermore, our results show that the true asymptotic
zero-temperature behaviour and the true low-temperature behaviour can
be seen only beyond the length scales currently studied with 
Monte-Carlo
simulations. 

The outline of this paper is as follows: In section \ref{simu} we
present our numerical results for the overlap distribution, the 
Binder
parameter, and the recursion of the couplings within MKA. In section
\ref{theory}, we give scaling arguments that yield  the
asymptotic exponents and the crossover behaviour seen in the
simulations. Section \ref{concl} summarizes and discusses the 
results.

\section{Numerical results}
\label{simu}

\begin{figure}
\centerline{
\epsfysize=0.2\columnwidth{\epsfbox{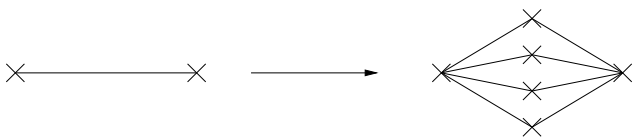}}}
\narrowtext{\caption{Construction of a hierarchical 
lattice.}\label{fig1}}
\end{figure}

The Migdal-Kadanoff approximation is a real-space
renormalization group the gives approximate recursion relations for
the various coupling constants.  Evaluating a thermodynamic quantity
in MKA in $d$ dimensions is equivalent to evaluating it on an
hierarchical lattice that is constructed iteratively by replacing 
each
bond by $2^d$ bonds, as indicated in Fig.~\ref{fig1}. The total 
number
of bonds after $I$ iterations is $2^{dI}$. $I=1$, the smallest
non-trivial system that can be studied, corresponds to a system 
linear
dimension $L=2$, $I=2$ corresponds to $L=4$, $I=3$ corresponds to
$L=8$ and so on. Note that the number of bonds on hierarchical 
lattice
after $I$ iterations is the same as the number of sites of a
$d$-dimensional lattice of size $L=2^I$.  Thermodynamic quantities 
are
then evaluated iteratively by tracing over the spins on the highest
level of the hierarchy, until the lowest level is reached and the
trace over the remaining two spins is calculated
\cite{southern77}. This procedure generates new effective couplings,
which have to be included in the recursion relations. The recursion
relation of the width $J(L)$ of the two-spin coupling is for
sufficiently many iterations and sufficiently low temperature given 
by
$J(L) \propto L^{\theta}$, with $\theta \simeq 0.26$ in MKA in three
dimensions (which is the only dimension studied in this paper).

\begin{figure}
\begin{center}
\epsfxsize=\columnwidth
\epsfbox{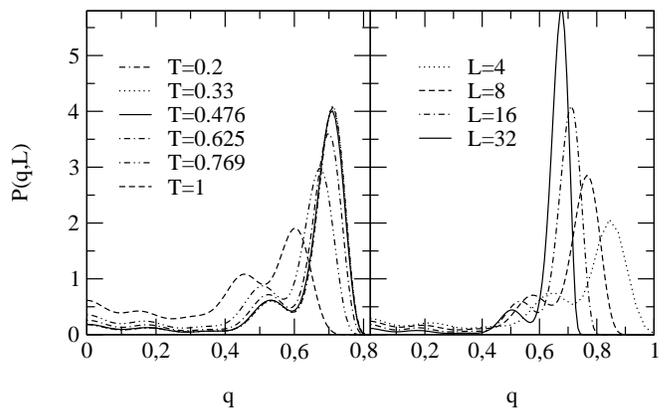}
\end{center}
\caption{The overlap distribution for the $\pm J$ spin glass in MKA
for $L=16$ and $T=1$, 0.769, $0.625$, 0.476, $0.33$, $0.2$ (left) and
for $T=0.33$ and $L=4$, 8, 16, 32 (right), all averaged over several 
thousand
samples.}
\label{fig2}
\end{figure}

We first evaluated the overlap distribution 
\begin{equation} 
P(q, L) = \left[\left<\delta \left(\sum_{\langle i j\rangle} 
\frac{{S_i^{(1)}S_i^{(2)} + S_j^{(1)}S_j^{(2)}}} 
{2N_L} - q
\right)\right>\right],
\label{p}
\end{equation}
between two identical replicas of the system, where the superscripts
$(1)$ and $(2)$ denote the two replicas of the system, $N_L$ is the
number of bonds of a system of size $L$, and $\langle ...\rangle$ and
$\left[...\right]$ denote the thermodynamic and disorder average
respectively. 
As discussed in \cite{ourprl}, the calculation of $P(q,L)$ is made
easier by first calculating its Fourier transform $F(y,L)$, which is
given by
\begin{equation}
F(y,L)=\left[\left< \exp\left(iy\sum_{\langle i j\rangle}
{(S_i^{(1)}S_i^{(2)}+S_j^{(1)}S_j^{(2)})\over {2N_L}}\right)
\right>\right] . \label{Fy}
\end{equation}
The recursion relations for $F(y,L)$ involve two-
and four-spin terms, and can easily be evaluated numerically 
because all
terms are now in an exponential. Having calculated $F(y,L)$, one 
can
then invert the Fourier transform to get $P(q,L)$. 
Figure \ref{fig2} shows our results for $L=16$ at
different temperatures, and for $T=0.33$ at different system sizes
respectively. Due to the ground state degeneracy, the overlap
distribution for fixed $L$ does not change below a temperature
for which most samples are in the ground state. With increasing 
system
size, the peaks in the overlap distribution become sharper, and the
probability of finding an overlap value near zero decreases,
indicating that one state and its spin-flipped counterpart dominate
the statistics. 

\begin{figure}
\centerline{ \epsfysize=0.6\columnwidth{\epsfbox{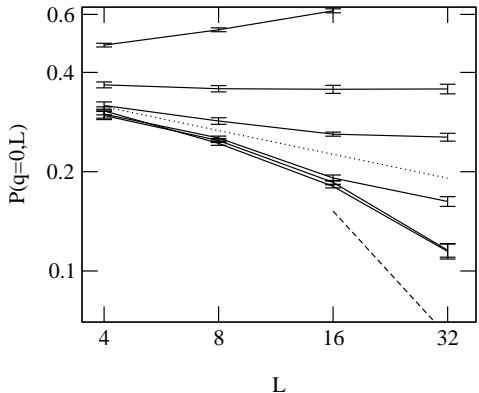}}}
\narrowtext{\caption{The probability density $P(q=0,L)$ for the $\pm
J$ spin glass in MKA for $T=1$, 0.769, 0.625, 0.476, 0.33, and 0.2
(from top to bottom). The critical temperature is $T_c=1.14$. The 
bars
indicate the standard deviation of the mean. The dotted line has the
droplet picture slope $-\theta=-0.26$, which must be for sufficiently
large $L$ the asymptotic slope of all shown curves. The dashed line
has the slope -1.26, which is the asymptotic slope expected for
sufficiently large $L$ at $T=0$.}\label{fig3}}
\end{figure}

Figure \ref{fig3} shows the probability
density $P(q=0,L)$ that the two replicas have zero overlap, at several
different temperatures. The two
curves for $T=0.33$ and $T=0.2$ are on top of each other, indicating
that at these sizes and temperatures, the system is in the ground
state with a probability close to 1. The $T=0.625$ curve coincides 
for
$L<16$ with the $T=0$ curve, but branches off for larger $L$ and
approaches the slope -0.26 expected from the droplet picture and seen
in a system with Gaussian distributed couplings (see
\cite{ourprl}). The $T=0.476$ curve seems to be affected by  ground
state effects for $L \le 16$, as it starts out close to the $T=0$ 
curve
and then has a negative slope which becomes flatter for larger $L$. A 
slope
flatter than that predicted by the droplet picture indicates an
influence of the critical point, as discussed in \cite{ourprl}. For
even larger $L$, the curve must become steeper again and approach the 
slope of the droplet picture. Even the $T=0.769$ curve appears to be
affected by the ground state degeneracy for $L \le 8$.  

From these numerical results, the asymptotic behaviour of the $T=0$
curve cannot be predicted. It seems unlikely that it becomes flatter
for larger $L$, implying that it is fundamentally different from the
droplet picture, which should govern the behaviour of a sufficiently
large system at low temperatures. The asymptotic slope of the $T=0$
curve and the crossover length scale at which a finite-temperature
curve branches off from it will be derived further below. 
Furthermore,
our results show that at system sizes smaller than around 16, the
asymptotic low-temperature behaviour is not visible for any
temperature, as the influence of the critical point and of the ground
state are too strong. 

\begin{figure}
\centerline{ \epsfysize=0.6\columnwidth{\epsfbox{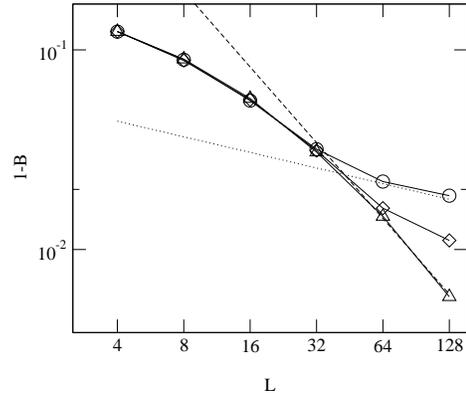}}}
\narrowtext{\caption{
The Binder parameter $B$ as function of system size for the $\pm
J$ spin glass in MKA for $T=0.33$, 0.2 and 0
(from top to bottom). The dotted line has a slope -0.26, the dashed 
line has the slope -1.26. All data points are averaged over 50000 
samples.
}\label{fig4}}
\end{figure}

In order to be able to study larger system sizes, we determined the
Binder parameter
\begin{equation}
B=\frac{3}{2}\left(1-\frac{[\langle q^4\rangle]}{3[\langle q^2 
\rangle]^2}
\right) , \label{binder}
\end{equation}
which can be obtained by differentiating Eq. (\ref{Fy}) with respect 
to $y$.
 This
is done by evaluating $F(y,L)$ for 3 small values of $y$. The
systematic error resulting from the finiteness of $y$ is found by
evaluating $F(y)$ for a few samples for many values of $y$, and by
extrapolating to $y \to 0$.  The Binder parameter  $B=0$ in the
high-temperature phase, and approaches 1 in the low-temperature phase
if the overlap distribution is trivial. Within the droplet picture, 
$1-B$ must
scale as $L^{-\theta}$ for sufficiently large $L$. Figure \ref{fig4}
shows our results. As for the overlap distribution, the $T=0$ curve 
is
much steeper than the limit slope expected from the droplet picture, 
and
the low-temperature curves branch off from it at a system size that 
is
larger for lower temperatures. In contrast to Figure  \ref{fig3},
system sizes can be studied that are large enough to see the
differences between the three curves for $T=0.33$, 0.2 and 0. The 
$T=0$
simulation was done by taking the trace only over those 
configurations
that contribute to the ground state, and by keeping track of the
degeneracies. 

\begin{figure}
\centerline{ \epsfysize=0.6\columnwidth{\epsfbox{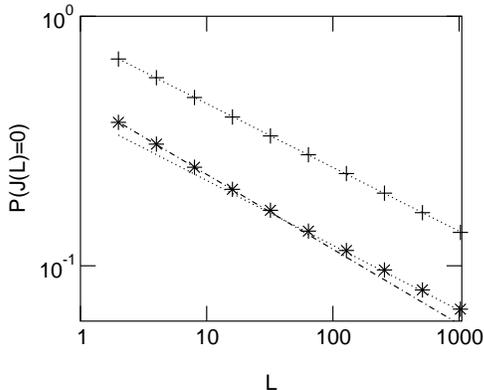}}}
\narrowtext{\caption{The probability of having a domain wall with no
energy cost in the ground state as function of the system size (* 
symbols).
For comparison, the data obtained for a Gaussian coupling 
distribution are also shown (+ symbols). The dotted lines have the 
slope -0.26, the dashed line has the slope -0.3.
}\label{fig5}}
\end{figure}

Next, we tried to understand the reasons for the steep decline of the
$T=0$ curve, and of its slow approach to the asymptotic slope. For
sufficiently large system sizes, the main contribution to $P(q=0,L)$
at $T=0$ must come from samples where a domain wall costs no energy. 
A
domain wall is introduced into the system by flipping one of the two
corner spins of the hierarchical lattice out of the ground state
orientation, and by determining the new ground state resulting with
this boundary condition. Samples where such a domain wall costs no
energy have zero effective coupling strength $J(L)=0$ at length scale
$L$, with $J(L)$ resulting from the recursion relation for the width
of the distribution of the couplings under the renormalization
procedure.  Figure \ref{fig5} shows the probability of having
$J(L)=0$ (or, equivalently, of having a domain wall with zero
energy cost) as function of the system size $L$. One can see that the
slope is identical to that expected from the droplet picture beyond
length scales $L=32$, and is only slightly steeper for smaller system
sizes. This indicates that the $\pm J$ model has a crossover length
around 32, which is not present in the model with Gaussian 
distributed
couplings, where the slope agrees with the droplet picture even for
the smallest system sizes.

\begin{figure}
\centerline{ \epsfysize=0.6\columnwidth{\epsfbox{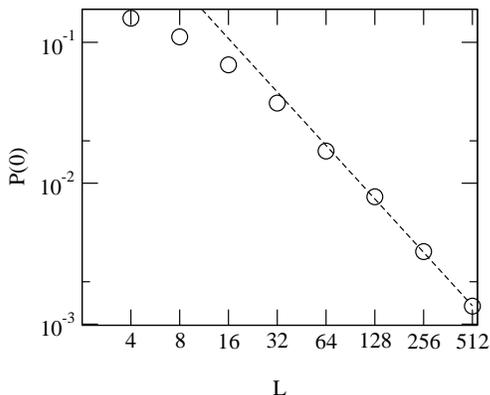}}}
\narrowtext{\caption{The probability that the states of two replicas
at $T=0$ have a different relative orientation of their corner
spins. The asymptotic slope, given by the dashed line, has the value
-1.26.  }\label{fig6}}
\end{figure}

Since Figure \ref{fig5} agrees with the droplet picture, it cannot
explain the steep decrease of $P(q=0,L)$, and $B$ at zero
temperature. For small system sizes, not only domain walls, but also
several small droplets can create a zero overlap, but this effect
should become irrelevant for sufficiently large system sizes. The only
remaining possibility is that while the probability for having a
domain wall of zero energy agrees with the droplet picture, the
weights of the two ground states with and without a domain wall differ
by a factor that increases with increasing system size. In order to
check this hypothesis, we evaluated the degeneracies of the two ground
states that are obtained by fixing the two corner spins in parallel
and antiparallel orientation respectively, and derived from this the
probability that two identical replicas of the system are in the two
different states. The result is shown in Figure \ref{fig6}. The curve
is steeper by $-1$ compared to that of the droplet picture, indicating
that entropy differences between ground states are the crucial factor
causing the deviation of the $T=0$ results from the droplet picture.
Figure \ref{fig6} shows the same slow approach towards asymptopia as
Figures \ref{fig3} and \ref{fig4}. We will attempt an explanation in
the next section which is devoted to a theoretical explanation of the
numerical findings.

\section{Scaling arguments}
\label{theory}

The main objective of the scaling theory presented here is to derive
the asymptotic slope of the $T=0$ curves, and to predict the 
crossover
length scale at which curves at finite $T$ branch off from the $T=0$
curve. As shown by our numerical data presented in the previous
section, entropy differences between ground state configurations that
differ by a domain wall play a crucial role. Let us therefore 
consider
a system that has a ground state for which a domain wall costs no
energy, and let us estimate the order of magnitude of the entropy
difference between the two ground states. One of the ground states is
obtained by fixing the two corner spins of the system (those with the
highest coordination number) in parallel orientation, and the other 
is
obtained by making them antiparallel. Contributions to the
entropy of each of these two states are made by droplet excitations
that cost no energy. (A droplet is a block of spins that are 
connected
to each other and that does not include one of the two corner spins,
and it may comprise just a single spin). By flipping several 
droplets,
one can thus get from every configuration contributing to one of the
two ground states (with fixed corner spins) to every other
configuration contributing to this state. The argument made in the
following is similar to the one made by Krzakala and Martin \cite{KM}
for a system with supposed RSB in the low-temperature phase. To each
configuration contributing to the first ground state (with parallel
corner spins), there exist configurations in the second ground state
(with antiparallel corner spins) that differ from it only by a domain
wall. This means that the two configurations can be transformed into
each other by flipping a coherent block of spins including the right
corner spin (assuming that the left corner spin is up in both 
states).
Now, all the possible zero-energy droplets in the first configuration
that do not touch the domain wall, are also zero-energy droplets in
the second configuration. Droplets that do not touch domain walls can
therefore make no contribution to the entropy difference between the
two ground states, because they occur in both of them. The entropy
difference between the two states results therefore from those
droplets that touch the domain wall. Now, the domain wall involves
$\propto L^{d_s}$ bonds, where $d_s$ is the fractal dimension of the
domain wall, and has the value $d_s=d-1$ in MKA. The average number 
of
droplets touching the domain wall can therefore be expected to be
$\propto L^{d_s}$ (assuming that the majority of droplets are small
and independent from each other), and the typical fluctuation
(measured over different samples, or over the two ground states) in
the number of droplets touching a domain wall can be expected to be
$\propto L^{d_s/2}$, which is identical to $L$ in MKA in three
dimensions. 

Now, the probability that two replicas have zero overlap is
proportional to the probability that a domain wall costs no energy,
$\propto L^{-\theta}$, multiplied by the probability that the
configurations of the two replicas have a different relative
orientation of the corner spins, $\propto L^{-d_s/2}$. This explains
the asymptotic slope of $-\theta-d_s/2 \simeq -1.26$ seen in Figure
\ref{fig6}, and expected for the zero-temperature $P(q=0,L)$ curve in
Figure \ref{fig3}. For the Binder parameter, Figure \ref{fig4}, we
expect the same asymptotic behaviour, $1-B \sim
L^{-\theta-d_s/2}$. The reason is that for large system sizes mainly
samples with zero-energy domain wall excitations show a considerable
difference between $\langle q^4 \rangle$ and $\langle q^2 \rangle^2$.

Next, let us discuss possible reasons why the asymptotic slope
$-\theta-d_s/2$ is approached so slowly in all our plots. Our scaling
argument is based on the assumption that the size distribution of the
droplets that touch the domain wall does not change much with the
system size. For small system sizes, the droplet size distribution
might be far from the asymptotic droplet size distribution, possibly
causing considerable deviations from asymptopia. This effect is
probably more severe in MKA than on a three-dimensional lattice,
because in MKA droplets consisting of a single spin with coordination
number two make no contribution to the entropy difference between the
two ground states. The reason is that flipping the domain wall
transforms each spin next to the domain wall with coordination number
2 that can be flipped without energy cost (and which therefore makes 
a
contribution to the entropy) into a spin that can be flipped only by
paying the energy 4, while every spin with coordination number two
along the domain wall that can be flipped only by paying energy, is
transformed into a spin that can be flipped without energy cost. The
numbers of the two classes of spins must therefore be equal, and the
entropy contribution due to spins with coordination number two that 
can
be flipped without energy cost is the same for both ground states.

Furthermore, we have assumed that the fluctuation in the numbers of
droplets touching the domain wall is given by the central limit
theorem, which is a good approximation only for sufficiently large
system sizes. Deviations from the numbers predicted by the central
limit theorem may be a further reason why the asymptotic slope
$-\theta-d_s/2$ is only visible for large system sizes.

Third, we have assumed that domain walls make the main contribution 
to
$P(q=0,L)$. This assumption is not correct for small system sizes,
where droplet excitations that do not involve the corner spin may 
also
add up to an overlap value of zero.

Finally, let us determine the crossover length scale beyond which the
droplet picture should become visible for small nonzero temperatures:
Within the droplet picture, we have $P(q=0,L) \sim TL^{-\theta}$,
while we have at zero temperature $P(q=0,L) \sim L^{-\theta-d_s/2}$. 
A
crossover between the two regimes occurs when the two quantities are 
equal,
i.e., when $L \sim T^{-d_s/2}$.

\section{Conclusions}
\label{concl}

In this paper, we have studied the $\pm J$ Ising spin glass within
MKA. We have found that the zero-temperature behaviour is
fundamentally different from that at low temperatures, due to entropy
differences between ground states.  Only for length scales larger 
than of the order
$T^{-d_s/2}$ does the expected droplet-picture behaviour become
visible. We have presented a scaling theory that predicts the
asymptotic scaling exponent $-\theta-d_s/2$ for the overlap
distribution at zero temperature, and we have shown from numerical
results as well as from analytical arguments that the approach to 
this
asymptotic scaling might be slow.

Our findings shed some light on recent Monte-Carlo simulations of the
three-dimensional $\pm J$ Ising spin glass. While the scaling
arguments by Krzakala and Martin \cite{KM} predict a decrease of
$P(q=0,L)$ at zero temperature at least with an exponent $-d_s/2$ (if
one assumes with them that the system shows RSB, implying 
$\theta=0$),
which lies somewhere between -1.1 and -1.3, the best Monte-Carlo
simulations find only a value around -0.9 \cite{PY,BHC}. Other 
Monte-Carlo
simulations giving a considerably smaller exponent probably do not
sample the ground state configurations with the appropriate weights
(see the comment by Marinari et al \cite{mprz} on the simulations by
Hatano and Gubernatis \cite{HG}, and the remarks by Palassini and
Young \cite{PY} on the simulations by Hartmann \cite{hart00}.)  Our
findings of a surprisingly slow approach to the correct asymptotic
scaling can reconcile the Monte-Carlo results with the predictions by
Krzakala and Martin, and also with our predictions based on the
droplet picture (where the asymptotic exponent is around -1.4 or
-1.5), which we believe to be the correct description of the
spin-glass phase. 

Our results in Figure \ref{fig3} show also that for not too low
temperatures the overlap distribution data may at first (for the
smallest $L$ values) be affected by the ground-state degeneracy (as
indicated by a slope that is initally steeper than for larger $L$),
then (for somewhat larger $L$) by the critical point (manifesting
itself in a pretty flat slope), and only for sufficiently large $L$
(which may be beyond the reach of Monte-Carlo simulations) the
correct asymptotic slope given by the droplet picture. Given such a
complicated behaviour, the predictions in \cite{HHD} and \cite{PY} 
for
the asymptotic behaviour of $P(q,L)$ based on small system sizes and
assuming simple scaling forms have no convincing basis.

We conclude that it is possible that the $\pm J$ Ising spin glass
in three dimensions and for system sizes smaller than approximately 
16 does not show the correct asymptotic scaling
behaviour at any value of the temperature.
It remains to be seen whether the Ising spin glass with a
Gaussian bond distribution has also finite-size effects which make it
impossible to see even at low temperatures the correct asymptotic
scaling behaviour for the system sizes presently used in computer
simulations.

\acknowledgements 
BD acknowledges support from the Deutsche Forschungsgemeinschaft, 
grant
number Dr300/2-1.

\end{multicols} 
\end{document}